# Teachers' Perspectives on the Use of AI Detection Tools: Insights from Ridge Regression Analysis

Vicky P. Vital[1], Francis F. Balahadia[2], Maria Anna D. Cruz[1], Dolores D. Mallari[1], Juvy C. Grume[1], Erika M. Pineda[1], Jordan L. Salenga[1], Lloyd D. Feliciano[1], John Paul P. Miranda[1*]

1. **Pampanga State University**, Pampanga, Philippines
2. **Laguna State Polytechnic University**, Laguna, Philippines

**\* Correspondence:**
John Paul P. Miranda, Pampanga State University, jppmiranda@pampangastateu.edu.ph

## ABSTRACT

This study explores the perceptions of 213 Filipino teachers toward AI detection tools in academic settings. It focuses on the factors that influence teachers' trust, concerns, and decision-making regarding these tools. The research investigates how teachers' trust in AI detection tools affects their perceptions of fairness and decision-making in evaluating student outputs. It also explores how concerns about AI tools and social norms influence the relationship between trust and decision-making. Ridge Regression analysis was used to examine the relationships between the predictors and the dependent variable. The results revealed that trust in AI detection tools is the most significant predictor of perceived fairness and decision-making among teachers. Concerns about AI tools and social norms have weaker effects on teachers' perceptions. The study emphasized critical role of trust in shaping teachers' perceptions of AI detection tools. Teachers who trust these tools are more likely to view them as fair and effective. In contrast, concerns and social norms have a limited influence on perceptions and decision-making. For recommendations, training and institutional guidelines should emphasize how these tools work, their limitations, and best practices for their use. Striking a balance between policy enforcement and educator support is essential for fostering trust in AI detection technologies. Encouraging experienced users to share insights through communities of practice could enhance the adoption and effective use of AI detection tools in educational settings.

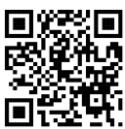

# INTRODUCTION

AI detection tools are software applications specifically designed to identify content generated by generative artificial intelligence (GenAI) (Perkins et al., 2024; Saqib & Zia, 2024). Their primary purpose is distinguishing authentic student work from outputs partially or entirely produced by artificial intelligence (AI) (Elkhatat et al., 2023; Perkins et al., 2024). Popular examples include Turnitin AI detection, CopyLeaks, Winston AI, and GPTZero. Despite their potential to enhance objectivity and efficiency in academic assessments, significant concerns persist regarding their reliability, accuracy, and vulnerability to biases (Cardona et al., 2023; Swiecki et al., 2022). These concerns are echoed in broader studies on AI adoption in education, such as (Jdaitawi et al., 2024), who found that perceived usefulness, ease of use, enjoyment, and social influence significantly influenced students' willingness to adopt AI tools in learning environments. Furthermore, reports from a university in the Philippines underscore inherent limitations, such as frequent inconsistencies and false positives which can undermine fairness and trust (De La Salle University, n.d.). This aligns with broader research indicating that AI detection tools cannot guarantee absolute accuracy, particularly given the rapid evolution of GenAI and the susceptibility of these tools to subtle edits made to AI-generated content (Elkhatat et al., 2023; Walters, 2023).

Within the Philippine educational context, however, research specifically addressing teachers' perceptions and their implications for the use of AI detection tools remains scarce. Teachers' perceptions significantly influence their trust and decision-making regarding the implementation of these tools in academic settings (Lyu et al., 2025). These perceptions may be shaped by their beliefs about the tools' reliability, their previous experiences with similar technologies, and biases arising from suspicions about student dishonesty (Huang et al., 2023; Lyu et al., 2025). Such biases might be especially salient in the Philippine educational landscape where cultural values place strong emphasis on academic integrity (Manarpiis & Prieto, 2023). Consequently, educators who already harbor suspicions toward particular students may either disproportionately rely on or disregard the results from AI detection tools which may impact their overall trust and implementation decisions (Merod, 2024; Zhai et al., 2024). Moreover, biases driven by the perceived convenience and ease of using these tools might further influence educators' consistent adoption (Yu, 2024).

Additionally, the roles of social norms and institutional support significantly shape educators' trust and practical decisions concerning AI detection tools (Lyu et al., 2025; Viberg et al., 2024). Social norms, characterized by collective attitudes, experiences, and peer expectations within academic institutions may include shared skepticism, encouragement, or hierarchical influence with the latter being particularly influential in Philippine educational institutions where senior educators and administrators significantly affect technology adoption decisions (Alegado, 2018; Maala & Lagos, 2022). Institutional support encompasses critical factors such as comprehensive training for teachers, appropriate technological infrastructure, and financial resources dedicated to technology implementation (Javier, 2022). Together, these institutional and normative elements either strengthen or diminish teachers' confidence in AI detection technologies

Considering the complexity of these interrelationships, this study addresses the identified research gap by examining how Filipino teachers' beliefs, biases, trust, institutional support, and social norms interconnect and influence their perceptions of AI detection tools. By employing path analysis, this research aims to provide a comprehensive understanding of these relationships within the Philippine educational context. Specifically, this study seeks to do the following:

     1.     To examine how teachers' trust in AI detection tools influences their perceptions of fairness and decision-making;

2. To explore how concerns about AI detection tools affect teachers' beliefs and trust in them; and
3. To assess the role of social norms and institutional influence in moderating the relationship between trust in AI tools and decision-making.

# METHOD

## Research Design and Participants

This study employed a descriptive cross-sectional research design. Path analysis was used to examine the relationships among variables. Data were collected through an online survey administered to teachers from various public and private schools in two provinces in the Philippines: Pampanga and Laguna. Participants were selected based on the following criteria: 1) at least one year of teaching experience, and 2) willingness to participate in the survey. The study included 213 respondents. Of these, 122 were married, and 78 were single. In terms of teaching levels, 116 respondents taught in basic education while 97 taught in higher education. The majority of respondents (9.6%, n = 193) worked full-time. Most held permanent positions (n = 164) and worked in public schools (n = 172). Additionally, 150 participants were licensed professional teachers. Regarding educational attainment, 102 respondents held a bachelor's degree, 73 held a master's degree, and 38 held a doctorate degree. A majority portion of respondents (58.7%, n = 125) reported no prior use or training in AI detection tools. Furthermore, 45.1% (n = 96) stated that their institutions did not provide access to such tools and 38% (n = 81) were unaware of whether their institutions offered or provided access to AI detection tools.

## Instrument

The instrument used in this study was developed based on a thorough literature review and operationalized into four constructs, each measured by multiple items. These constructs reflect teachers' attitudes and behaviors toward AI detection tools, including Perceived Effectiveness and Fairness (PEF), Concerns and Peer Influence (CPI), Social Norms and Institutional Influence (SNI), and Trust and Decision-Making (TDM). Each construct assesses different dimensions of teachers' interactions with AI tools, as outlined in Table 1. The instrument underwent validation by two educational technology professors, and reliability testing showed that Cronbach's alpha values for all constructs exceeded the required threshold of .70 which confirmed its internal consistency and reliability. This validated tool provides a comprehensive framework for investigating teachers' perspectives on AI detection tools in academic settings.

**Table 1.**
*Constructs, definitions, and basis*

| Construct | Definition | Basis |
|---|---|---|
| PEF | Measures the extent to which teachers perceive AI detection tools as fair and effective in evaluating student outputs. | (Chai et al., 2024; Memarian & Doleck, 2023) |
| CPI | Assesses the degree to which concerns about AI tools and peer influence affect teachers' perceptions and decision-making regarding AI tools. | (Fan & Zhao, 2023; McGehee, 2024; Nazaretsky et al., 2022; Velli & Zafiropoulos, 2024) |
| SNI | Captures the extent to which social norms, peer behaviors, and institutional policies influence teachers' perceptions and use of AI tools. | (Bakhadirov et al., 2024; Nazaretsky et al., 2022; Ofosu-Ampong, 2024; Velli & Zafiropoulos, 2024) |
| TDM | Evaluates the degree to which teachers' trust in AI tools influences their decision-making regarding student outputs. | (Nazaretsky et al., 2022; Viberg et al., 2024) |

## Data Collection Procedure and Ethical Consideration

This study utilized an online survey conducted between January and February 2025 to explore teachers' perceptions of AI detection tools in educational settings. The online survey method was chosen to ensure efficient and widespread data collection from educational practitioners across diverse contexts. Participants were recruited through institutional channels, including administrators, deans, and department chairpersons, as well as through networks of preservice teachers. Prior to participation, the research objectives were clearly explained to all potential respondents. Informed consent was obtained implicitly through their voluntary participation in the survey. To uphold ethical standards, strict confidentiality measures were implemented. All responses were anonymized and data were aggregated in the analysis to ensure that no individual could be identified. These procedures align with established ethical guidelines for research, safeguarding participant privacy, and integrity throughout the study.

### Statistical Analysis

Descriptive statistics were calculated for the four factors: PEF, CPI, SNI, and TDM. These statistics included the mean ($\bar{x}$) and standard deviation (SD) for each factor to provide an overview of the sample's distribution. A correlation matrix was generated to assess the relationships between the factors and guide the selection of the appropriate analysis. A Ridge Regression analysis was performed to investigate the relationships between the predictors (CPI, SNI, and TDM) and the dependent variable (PEF). Ridge Regression was chosen for its ability to handle complex relationships between variables by applying L2 regularization which stabilizes estimates by penalizing the size of the coefficients and improves the model's generalizability. The model was standardized to ensure comparability of the coefficients across predictors. All computations were made using Python in the Jupyter Notebook environment.

## RESULTS AND DISCUSSION

### Descriptive Summary

The overall descriptive summary of the four constructs is presented in Table 2. For PEF, teachers rated the usefulness of AI detection tools for evaluating student outputs ($\bar{x}$ = 3.73, SD = 1.03). Teachers believed that AI detection tools enhance fairness in assessing student outputs ($\bar{x}$ = 3.69, SD = 1.01). Teachers considered AI detection tools unnecessary for evaluating student outputs ($\bar{x}$ = 2.76, SD = 1.09). The overall mean for PEF was 3.44 (SD = .60). For CPI, teachers' trust in the accuracy of AI detection tool results ($\bar{x}$ = 3.54, SD = .74) was moderately high. Teachers expressed confidence in relying on AI tools for academic decisions ($\bar{x}$ = 3.88, SD = .95). Teachers were concerned that AI detection tools may not provide transparent justifications for their results ($\bar{x}$ = 3.61, SD = .93). The likelihood of using AI tools when suspecting student use of GenAI was ($\bar{x}$ = 3.76, SD = .84). The overall mean for CPI was $\bar{x}$ = 3.68 (SD = .58).

For SNI, teachers observed colleagues using AI detection tools to assess student outputs ($\bar{x}$ = 3.06, SD = .88). Teachers felt encouraged by peers to use AI detection tools ($\bar{x}$ = 3.14, SD = .86). Teachers' decision-making was influenced by colleagues' practices ($\bar{x}$ = 3.49, SD = .86). The overall mean for SNI was ($\bar{x}$ = 3.31, SD = .66). For TDM, teachers trusted AI tools' ability to accurately identify AI-generated content in student outputs ($\bar{x}$ = 3.62, SD = 1.00). Teachers also expressed confidence in relying on AI tools for academic decisions ($\bar{x}$ = 3.38, SD = .78). Teachers felt confident in using AI detection tools for academic decisions ($\bar{x}$ = 3.23, SD = .82). Teachers were more likely to use AI tools when suspecting GenAI usage ($\bar{x}$ = 3.76, SD = .84). The overall mean for TDM was 3.49 (SD = .65).

**Table 2.**
*Descriptive summary*

| Construct | $\bar{x}$ | SD |
|---|---|---|
| PEF | 3.44 | .60 |
| CPI | 3.68 | .58 |
| SNI | 3.31 | .66 |
| TDM | 3.49 | .65 |

Table 3 shows that PEF and TDM have a strong positive correlation (*r* = .70). PEF and show a moderate positive correlation (*r* = .49). The correlation between CPI and TDM is moderate (*r* = .37). The relationship between SNI and TDM is also moderate (*r* = .66).

**Table 3.**
*Correlation analysis*

| Construct | PEF | CPI | SNI |
|---|---|---|---|
| CPI | **.35** | | |
| SNI | .49 | **.25** | |
| TDM | .70 | .37 | **.66** |

### Trust in AI Detection Tools and Perceived Fairness

The results indicated that TDM had the strongest influence on PEF ($β$ = .623) which means that when teachers trust AI detection tools, they are more likely to perceive them as fair and reliable. This is aligned with previous studies which showed that faculty who view AI tools as valuable and beneficial are more inclined to use them, whereas skepticism about accuracy reduces trust (Chai et al., 2024). Similarly, (Jdaitawi et al., 2024) supported the notion that positive beliefs enhance trust and perceived fairness. In addition, studies have reported that trust in AI was a significant predictor of perceived usefulness which may reinforce that educators who trust AI tend to see it as a fair assessment tool (Prothero, 2024).

Moreover, Zawacki-Richter et al. (2019) emphasized that transparency, explainability, and consistent performance significantly impact educators' trust in AI systems. Teachers who understand the decision-making processes behind AI evaluations are more likely to trust and rely on these tools. However, concerns about potential biases and inaccuracies remain a challenge (Ali et al., 2024; Zhai et al., 2024), as highlighted by a Pew Research Center survey, which found that many K-12 teachers believe AI tools may do more harm than good (Lin, 2024). Addressing these concerns through increased transparency and training can foster trust and enhance perceived fairness.

### Concerns About AI Detection Tools and Their Influence on Trust

The results indicated that CPI had a weaker but still positive relationship with PEF ($β$ = .107). Research highlighted that while some teachers perceive AI detection tools as efficient and objective, others remain skeptical due to concerns about false positives, false negatives, and biases in detection (Giray et al., 2025). Moreover, studies believed that plagiarism-detection software was valuable but also acknowledged that AI tools could be manipulated which raises doubts about its fairness (Arabyat et al., 2022; Giray et al., 2025; Madhu et al., 2023). Matek et al. (2017) similarly emphasized that static, rigid, and opaque systems limit user trust and effectiveness. Recent analyses of AI text detectors have further amplified these concerns. Studies also found that these tools exhibit high false-positive rates particularly misidentifying non-native English writers' work as AI-generated at disproportionate rates (Giray et al., 2025). Even OpenAI discontinued its own detector after acknowledging its unreliability (Arasa, 2023). Such findings highlighted a critical issue in AI fairness: inaccurate tools can unfairly accuse students of misconduct which potentially undermine trust among educators (Giray et al., 2025).

### Social Norms and Institutional Influence in AI Decision-Making

The results shows that SNI had the weakest effect on PEF (β = 058) and indicates that while institutional support matters, it is not as influential as trust or concerns about AI effectiveness. Research consistently shows that formal support (such as AI policies and training) can reassure teachers about a tool's legitimacy (Chan, 2023; Nazaretsky et al., 2022). For example, a 2024 study found that teachers in schools with AI usage guidelines and leadership encouragement were more likely to use AI detection tools (Bakhadirov et al., 2024). However, institutional pressure without proper support can create resistance (Elstad & Eriksen, 2024). Some educators reported needing clearer guidelines and training on handling AI-generated content before fully trusting these tools (Cheng & Wang, 2023). A 2023 study found that mandates alone do not convince teachers of a tool's fairness (Arvin et al., 2023) and that resources such as workshops and technical support are necessary to improve perceptions (Arvin et al., 2023).

Peer influence also plays a role in shaping teachers' acceptance of AI detection tools, though its impact appears to be modest compared to personal beliefs (Velli & Zafiropoulos, 2024). While peer norms do matter, individual attitudes and self-efficacy appear more decisive in determining AI adoption (Shao et al., 2025). Studies found that faculty members' personal attitudes and moral obligation, rather than social influence or colleagues' use, are more influential in their adoption of plagiarism-detection software (Arabyat et al., 2022). However, other research has documented stronger peer effects. Bakhadirov et al. (2024) found that teachers who observed their colleagues frequently using AI tools were significantly more likely to adopt them themselves. Their study suggests that while direct peer pressure may not be highly influential, seeing successful use by colleagues legitimizes AI tools and increases confidence in their fairness.

## CONCLUSION AND RECOMMENDATIONS

This study demonstrated that trust functions as the primary determinant of faculty perceptions regarding AI detection tools, specifically concerning their effectiveness and fairness. Trust in AI tools is the strongest predictor of positive perceptions and has a greater influence than concerns about misuse or social norms. For educational institutions implementing these technologies, building and maintaining trust must be the central policy objective. Institutions should require independent evaluations of AI detection tools for accuracy and bias before adoption. Vendor transparency should be mandatory, including clear reporting on algorithmic processes, error rates, and system limitations. These criteria should be written into procurement policies. Institutions should also provide structured training for faculty, along with user guides and access to technical support, to reduce uncertainty and support responsible use.

To increase adoption, institutions must involve faculty in implementation efforts. Encouraging experienced users to lead peer workshops or contribute to communities of practice can strengthen trust through shared knowledge. Institutions should also create formal feedback systems such as faculty surveys and advisory committees to monitor tool performance and guide policy adjustments. Policies must avoid overly rigid enforcement and instead focus on support and collaboration. This includes setting clear expectations for use to ensure that detection results are used fairly and protects educators and students from unjust consequences. Institutions should promote shared decision-making when adopting AI systems. Future research should examine how specific policy interventions affect long-term trust and use in diverse academic environments.